\documentclass[floats, prd, eqnum, showpacs, nofootinbib, 
twocolumn, 
eqsecnum]{revtex4-1}

\usepackage{color,graphicx}
\usepackage{amsfonts}
\usepackage{amssymb}
\usepackage{comment}
\begin{document}

\title{Gravitational lensing in the Simpson-Visser black-bounce spacetime in a strong deflection limit}
\author{Naoki Tsukamoto${}^{1}$}\email{tsukamoto@rikkyo.ac.jp}
\affiliation{
${}^{1}$Department of General Science and Education, National Institute of Technology, Hachinohe College, Aomori 039-1192, Japan \\
}

\begin{abstract}
A Simpson-Visser spacetime has two nonnegative parameters $a$ and $m$ and its metric is correspond with
(i) a Schwarzschild metric for $a=0$ and $m\neq0$,
(ii) a regular black hole metric for $a<2m$, 
(iii) a one-way traversable wormhole metric for $a=2m$,
(vi) a two-way traversable wormhole metric for $a>2m$, and 
(v) an Ellis-Bronnikov wormhole metric for $a\neq0$ and $m=0$. 
The spacetime is one of the most useful spacetimes 
for the purpose
of comprehensively understanding gravitational lensing of light rays reflected by a photon sphere of black holes and wormholes.
We have investigated gravitational lensing in the Simpson-Visser spacetime in a strong deflection limit in all the nonnegative parameters of $a$ and $m$.
In a case of $a=3m$, two photon spheres and an antiphoton sphere at the throat degenerate into a marginally unstable photon sphere.
The deflection angle of the light rays reflected by the marginally unstable photon sphere at the throat diverges nonlogarithmically in the strong deflection limit.
\end{abstract}

\maketitle

\section{Introduction}
Recently, the detections of gravitational waves emitted by binary black holes 
and of the shadow of a supermassive black hole candidate at center of a giant elliptical galaxy M87 have been reported by the LIGO and VIRGO Collaborations~\cite{Abbott:2016blz}
and by Event Horizon Telescope Collaboration~\cite{Akiyama:2019cqa}, respectively. 
The phenomena in a strong gravitational field near compact objects can be more important in general relativity and astrophysics. 

Static and spherically symmetric compact objects with a strong gravitational field in general relativity have unstable (stable) circular photon orbits called photon spheres (antiphoton spheres)~\cite{Claudel:2000yi,Perlick_2004_Living_Rev}
and its property, such as the upper bound of the radius~\cite{Hod:2017xkz} and the number~\cite{Hod:2017zpi}, has been studied.
The relations between (anti)photon spheres and the photon absorption cross section~\cite{Sanchez:1977si,Decanini:2010fz,Wei:2011zw}
quasinormal modes~\cite{Press:1971wr,Goebel_1972,Stefanov:2010xz,Raffaelli:2014ola},
a centrifugal force and gyroscopic precession~\cite{Abramowicz_Prasanna_1990,Abramowicz:1990cb,Allen:1990ci,Hasse_Perlick_2002},
Bondi's sonic horizon of a radial fluid~\cite{Mach:2013gia,Chaverra:2015bya,Cvetic:2016bxi,Koga:2016jjq,Koga:2018ybs,Koga:2019teu},
stability of thin-shell wormholes~\cite{Barcelo:2000ta,Koga:2020gqd}, 
and an apparent shape during a collapsing star to be a black hole~\cite{Ames_1968,Synge:1966okc,Yoshino:2019qsh}
have been investigated.
Extensions and alternatives of the (anti)photon spheres to low symmetry have also been investigated~\cite{Claudel:2000yi,Koga:2019uqd,Cunha:2017eoe,Gibbons:2016isj,Shiromizu:2017ego,Yoshino:2017gqv,Galtsov:2019bty,Galtsov:2019fzq,Siino:2019vxh,Yoshino:2019dty,Cao:2019vlu,Yoshino:2019mqw,Lee:2020pre} 
and there is concerned that stable photon rings of compact objects lead to instability caused by the slow decay of linear waves~\cite{Keir:2014oka,Cardoso:2014sna,Cunha:2017qtt}.

Gravitational lensing has been studied not only in a weak gravitational field~\cite{Schneider_Ehlers_Falco_1992,Schneider_Kochanek_Wambsganss_2006} but also in a strong gravitational field.  
In 1931, Hagihara pointed out that light rays are strongly deflected by the photon sphere and that an observer will see the light rays coming from all the directions of our universe~\cite{Hagihara_1931}.  
Infinite number of the dim images of the light rays reflected by the photon sphere have been revived by many authors~\cite{Darwin_1959,Atkinson_1965,Luminet_1979,Ohanian_1987,Nemiroff_1993,
Frittelli_Kling_Newman_2000,Virbhadra_Ellis_2000,Bozza_Capozziello_Iovane_Scarpetta_2001,Bozza:2002zj,Perlick:2003vg,Nandi:2006ds,Virbhadra:2008ws,Bozza_2010,Tsukamoto:2016zdu,
Shaikh:2019itn,Shaikh:2019jfr,Wielgus:2020uqz,Paul:2020ufc,Tsukamoto:2020uay,Tsukamoto:2020iez}. 
The dim images are named relativistic images in Ref.~\cite{Virbhadra_Ellis_2000}.

The deflection angle of a light scattered by the photon sphere
in a general asymptotically flat, static and spherically symmetric spacetime
in a strong deflection limit $b \rightarrow b_{\mathrm{m}}$, 
where $b$ is the impact parameter of the light and $b_{\mathrm{m}}$ is the critical impact parameter,
is expressed by
\begin{eqnarray}\label{eq:def0}
\alpha&=&-\bar{a} \log \left( \frac{b}{b_{\mathrm{m}}}-1 \right) + \bar{b} \nonumber\\
&&+ O\left( \left( \frac{b}{b_{\mathrm{m}}}-1 \right)  \log \left( \frac{b}{b_{\mathrm{m}}}-1 \right) \right),
\end{eqnarray}
where $\bar{a}$ and $\bar{b}$ are parameters, and its application to a lens equation has
been investigated by Bozza~\cite{Bozza:2002zj} and the formalism has been extend by many authors~\cite{Bozza:2002af,Eiroa:2002mk,Petters:2002fa,
Eiroa:2003jf,Bozza:2004kq,Bozza:2005tg,Bozza:2006sn,Bozza:2006nm,Iyer:2006cn,Bozza:2007gt,Tsukamoto:2016qro,Ishihara:2016sfv,Tsukamoto:2016oca,
Tsukamoto:2016zdu,Tsukamoto:2016jzh,Tsukamoto:2017edq,Shaikh:2018oul,Shaikh:2019itn,Paul:2020ufc,Tsukamoto:2020uay,Tsukamoto:2020iez}~\footnote{In 
Ref.~\cite{Bozza:2002zj}, the subleading term of Eq.~(\ref{eq:def0}) has been considered as $O \left( b-b_{\mathrm{m}} \right)$, However, it should be read as $O\left( \left( \frac{b}{b_{\mathrm{m}}}-1 \right)  \log \left( \frac{b}{b_{\mathrm{m}}}-1 \right) \right)$
as discussed in Refs.~\cite{Iyer:2006cn,Tsukamoto:2016qro,Tsukamoto:2016jzh}.}.
Recently, deflection angles in the strong deflection limit with some different forms from Eq.~(\ref{eq:def0}) 
such as $\alpha=\bar{a}_* \left(b/b_{\mathrm{m}}-1\right)^{-1/6} +\bar{b}_* +O\left( \left( b/b_{\mathrm{m}}-1\right)^{1/6} \right)$, 
where $\bar{a}_*$ and $\bar{b}_*$ are constant, have been reported in Refs.~\cite{Tsukamoto:2020iez,Chiba:2017nml} 
when an antiphoton sphere and a photon sphere degenerate to a marginally unstable photon sphere.
These recent studies on the strong deflection limit analysis show that Bozza's standard method~\cite{Bozza:2002zj} 
and alternative methods~\cite{Tsukamoto:2016jzh} do not always work in the case of all parameters of spacetimes.
Thus, we have to choose carefully appropriate coordinates, variable $z$, and methods of the analysis.

Wormholes are hypothetical objects permitted as a solution of Einstein equations with nontrivial topology~\cite{Visser_1995,Morris_Thorne_1988}
and they do not have an event horizon but they can have photon spheres and antiphoton spheres.
It is known that any static and spherically symmetric wormhole violates energy conditions on a throat at least 
if we assume general relativity~\cite{Morris_Thorne_1988}.
Gravitational lensing of light rays passing through 
the throat or passing near the photon sphere~\cite{Chetouani_Clement_1984,Perlick:2003vg,Nandi:2006ds,Muller:2008zza,Tsukamoto_Harada_Yajima_2012,Perlick:2014zwa,Tsukamoto:2016qro,
Tsukamoto:2016zdu,Nandi:2016uzg,Tsukamoto:2017edq,Shaikh:2018oul,Bhattacharya:2019kkb,Shaikh:2019jfr},
the visualizations of wormholes~\cite{Muller_2004,James:2015ima}, and
shadows in an accretion gas~\cite{Ohgami:2015nra,Ohgami:2016iqm,Paul:2019trt}, 
wave optics~\cite{Nambu:2019sqn},
and gravitational waves~\cite{Cardoso:2016rao}
in wormhole spacetimes have been investigated.

Recently, Simpson and Visser have suggested a metric which can correspond with
a Schwarzschild metric ($a=0$ and $m\neq0$), a regular black hole metric ($a<2m$), and a wormhole metric ($a\geq 2m$) including an Ellis-Bronnikov wormhole metric ($a\neq0$ and $m=0$),
where $a$ and $m$ are parameters of the spacetime~\cite{Simpson:2018tsi}.
Assuming general relativity, the energy conditions must be violated as shown in~\cite{Simpson:2018tsi}. 
This is not surprising because the Simpson-Visser metric includes the wormhole metric as a special case.
We notice that Simpson and Visser have disregarded the photon sphere of the wormhole with $a>3m$.
Its gravitational lensing in the strong deflection limit for $a<3m$~\cite{Nascimento:2020ime}
and the one under a weak-field approximation~\cite{Nascimento:2020ime,Ovgun:2020yuv} have been studied.
New examples of spacetimes similar to the Simpson-Visser spacetime have been proposed in Ref.~\cite{Lobo:2020ffi}.

In this paper, we investigate the gravitational lensing in the strong deflection limit in the Simpson-Visser spacetime in all the cases of the non-negative parameters~$a$ and $m$.
We show that the observation of the black hole shadow~\cite{Akiyama:2019cqa} does not reject the wormhole with $a>3m$ 
which is disregarded in~\cite{Simpson:2018tsi,Nascimento:2020ime}.
We also show that two photon spheres and one antiphoton sphere degenerate into a marginally unstable photon sphere at a wormhole throat for $a=3m$
and that the deflection angle in the strong deflection limit becomes
\begin{equation}\label{eq:def02}
\alpha(b)=\frac{\bar{c}}{\left(\frac{b}{b_{\mathrm{m}}}-1\right)^\frac{1}{4} } +\bar{d}, 
\end{equation}
where $\bar{c}$ and $\bar{d}$ are constant while it has the form of Eq.~(\ref{eq:def0}) for $a\neq3m$.

This paper is organized as follows. 
In Sec.~II, we review the Simpson-Visser spacetime and the deflection angle of the light ray. 
In Secs.~III and IV, we investigate the deflection angle and observables in the strong deflection limit. 
Gravitational lensing under a weak-field approximation is shortly reviewed in Sec.~V,  and this paper is concluded in Sec. VI.
In this paper we use the units in which a light speed and Newton's constant are unity.

\section{Simpson-Visser spacetime}
The Simpson-Visser spacetime is described by a line element, in Buchdahl coordinates \cite{Finch:1998,Boonserm:2007zm}, 
$-\infty <t < \infty$, $-\infty <r < \infty$, $0 \leq  \vartheta \leq \pi$, and $0 \leq  \varphi < 2\pi$,
\begin{eqnarray}\label{eq:line_element} 
ds^2
=-A(r)dt^2 +B(r)dr^2 + C(r)(d\vartheta^2+\sin^2\vartheta d\varphi^2), \nonumber\\
\end{eqnarray}
where $A(r)$, $B(r)$, and $C(r)$ are given by
\begin{eqnarray}
A(r)&=&\frac{1}{B(r)}\equiv 1- \frac{2m}{\sqrt{r^2+a^2}}, \nonumber\\
C(r)&\equiv&r^2+a^2
\end{eqnarray}
and where $m$ and $a$ are non-negative parameters.
It is 
(i) a Schwarzschild metric if $a=0$ and $m\neq0$,
(ii) a regular black hole metric if $a<2m$, 
(iii) a one-way traversable wormhole metric with a null throat if $a=2m$,
(vi) a traversable wormhole metric with a two-way throat at $r=0$ if $a>2m$, and 
(v) an Ellis-Bronnikov metric if $a\neq0$ and $m=0$. 
On this paper, we use not only the radial coordinate $r$ but also a standard radial coordinate $\rho \equiv \sqrt{r^2+a^2}$ which is related to the surface area $4\pi \rho^2$ of a two-dimensional sphere.
See the end of this section for the standard radial coordinate~$\rho$.
There are time-translational and axial Killing vectors 
$t^\mu \partial_\mu=  \partial_t$ and $\varphi^\mu \partial_\mu = \partial_\varphi$
because of stationarity and axisymmetric symmetry of the spacetime, respectively.
Without loss of generality, we concentrate on $\vartheta=\pi/2$ and $r\geq 0$.

The trajectory of the light ray is described by $k^\mu k_\mu=0$, 
where $k^\mu \equiv \dot{x}^\mu$ is the wave number of the light 
and where the dot denotes the differentiation with respect to an affine parameter along the trajectory.
The equation of the trajectory of the light is written as 
\begin{eqnarray}\label{eq:trajectory} 
-A(r)\dot{t}^2+B(r)\dot{r}^2+C(r)\dot{\varphi}^2=0.
\end{eqnarray}
We consider a light ray comes from spatial infinity, it is reflected by a black hole or wormhole at the closest distance $r=r_0$, it goes back to spatial infinity. 
At the closest distance $r=r_0$, Eq.~(\ref{eq:trajectory}) becomes
\begin{eqnarray}\label{eq:trajectory1} 
A_0\dot{t}_0^2=C_0\dot{\varphi}_0^2.
\end{eqnarray}
Here and hereafter, functions with subscript $0$ denote the functions at $r=r_0$. 
From Eq.~(\ref{eq:trajectory1}), an impact parameter $b$ is expressed by
\begin{eqnarray}\label{eq:b_r0} 
b(r_0)\equiv \frac{L}{E}=\frac{C_0 \dot{\varphi}_0}{A_0 \dot{t}_0}=\pm \sqrt{\frac{C_0}{A_0}},
\end{eqnarray}
where $E\equiv -g_{\mu \nu}t^{\mu}k^{\nu}$ and $L\equiv g_{\mu \nu}\varphi^{\mu}k^{\nu}$ are the conserved energy and angular momentum of the light, respectively, 
and they are constant along the trajectory.

Equation~(\ref{eq:trajectory}) is rewritten as
\begin{eqnarray}
\dot{r}^2+V(r)=0,
\end{eqnarray}
where $V(r)$ is an effective potential defined as
\begin{eqnarray}
V(r)\equiv -\frac{L^2 F(r)}{B(r)C(r)}
=\frac{L^2}{C(r)B(r)}-E^2
\end{eqnarray}
where $F(r)$ is given by
\begin{eqnarray}
F(r)\equiv \frac{C(r)}{A(r)b^2}-1.
\end{eqnarray}
The light ray can exist in a region for $V(r)\leq  0$.
The first, second, third, and forth derivatives of $V(r)$ with respect to the radial coordinate $r$ are given by
\begin{eqnarray}
V^{\prime}=\frac{2L^2 r \left( 3m -\sqrt{a^2+r^2} \right)}{\left(a^2+r^2 \right)^\frac{5}{2}},
\end{eqnarray}
\begin{eqnarray}
V^{\prime \prime }&=&\frac{2L^2}{\left(a^2+r^2 \right)^\frac{7}{2}} \left[ -3r^2\left( 4m-\sqrt{a^2+r^2} \right)  \right. \nonumber\\
&&\left. +a^2\left( 3m-\sqrt{a^2+r^2} \right) \right] , 
\end{eqnarray}
\begin{eqnarray}
V^{\prime \prime \prime }
&=&\frac{6L^2r}{\left(a^2+r^2 \right)^\frac{9}{2}} \left[ 4r^2 \left( 5m-\sqrt{a^2+r^2} \right) \right. \nonumber\\
&&\left. +a^2 \left( -15m+4\sqrt{a^2+r^2} \right) \right] , 
\end{eqnarray}
and
\begin{eqnarray}
V^{\prime \prime \prime \prime }
&=&\frac{6L^2}{\left(a^2+r^2 \right)^\frac{11}{2}}  \left[ 20a^2r^2 \left( 9m-2\sqrt{a^2+r^2} \right)   \right. \nonumber\\
&&+20 r^4 \left( -6m+\sqrt{a^2+r^2} \right) \nonumber\\
&&\left. +a^4  \left( -15m + 4\sqrt{a^2+r^2} \right) \right], 
\end{eqnarray}
respectively.
We name a stable (unstable) circular light orbit which satisfies $V=V^{\prime}=0$ and $V^{\prime \prime}<0$ ($V^{\prime \prime}>0$)
photon sphere (antiphoton sphere).
Let $r_\mathrm{m}$ the radius of the outermost circular light orbit which satisfies $V_\mathrm{m}=V^{\prime}_\mathrm{m}=0$.
Here and hereafter, functions with the subscript $m$ denotes the functions at the outermost circular light orbit.
The light ray with the impact parameter $b<b_\mathrm{m}$, where 
$b_\mathrm{m} = b(r_\mathrm{m})$ is the critical impact parameter, falls into the black hole or the wormhole 
while the light ray with the impact parameter $b>b_\mathrm{m}$ is scattered by the black hole or the wormhole. 
Figure~1 shows a dimensionless effective potential $V(r)/E^2$ for the light ray with the critical impact parameter $b=b_\mathrm{m}$.
In Refs.~\cite{Simpson:2018tsi,Nascimento:2020ime}, a circular light orbit at $r=0$ has been disregarded.
It is inside the event horizon for $a<2m$ 
while it is an antiphoton sphere for $2m\leq a < 3m$,
which is the stable circular light orbit with the impact parameter $b=\sqrt{\frac{a^3}{a-2m}}$, 
satisfies~$V(0)=V^{\prime}(0)=0$ and $V^{\prime \prime }(0)>0$ 
and it is coincident with the wormhole throat at $r=0$.
However, we cannot disregard the gravitational lensing by the circular light orbit at $r=0$ for $3m \leq a$ 
since it becomes the (marginally unstable) photon sphere as shown in Fig.~1.
We concentrate on the scattering case $b>b_\mathrm{m}$.
We name $r_0 \rightarrow r_\mathrm{m}$ or $b \rightarrow b_\mathrm{m}$ strong deflection limit. 
\begin{figure}[htbp]
\begin{center}
\includegraphics[width=87mm]{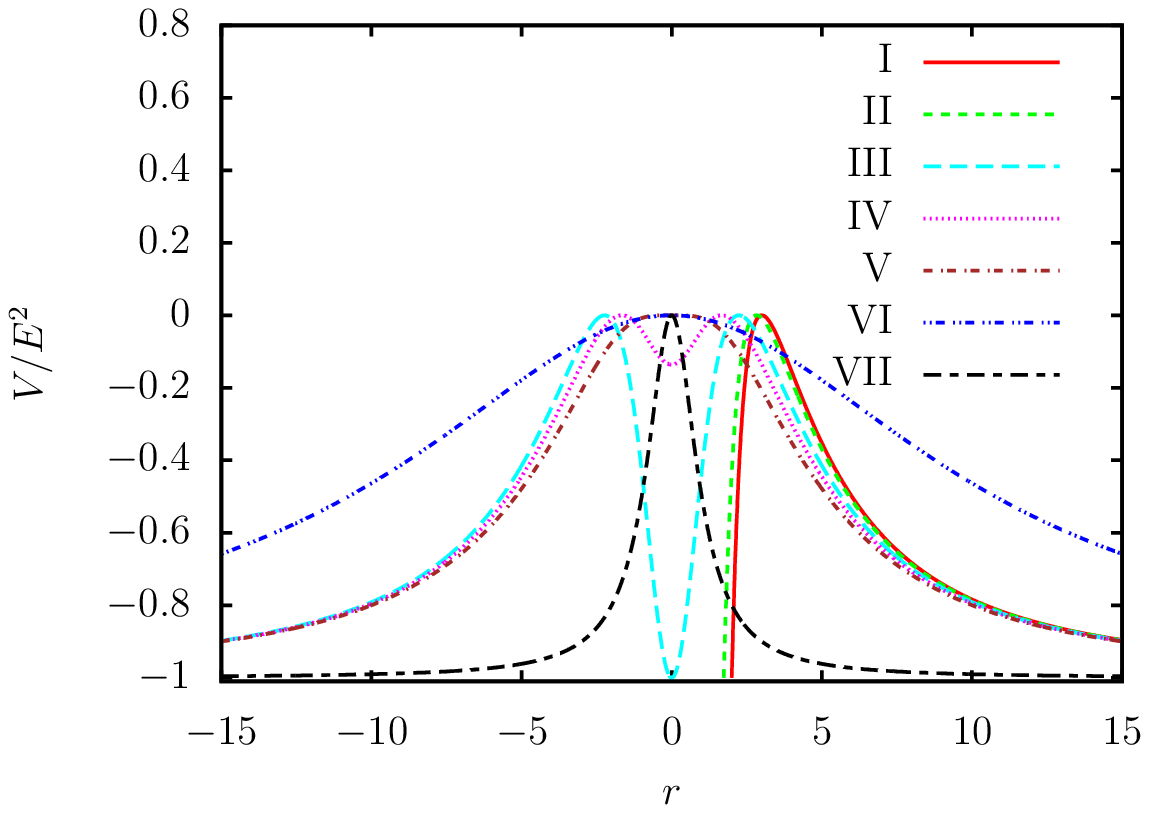}
\end{center}
\caption{A dimensionless effective potential $V/E^2$ of a light ray with $b=b_\mathrm{m}$ as a function of the radial coordinate~$r$.
Solid~(red), dashed~(green), long-dashed~(cyan), dotted~(magenta), dotted-dashed~(brown), 
double-dotted-dashed~(blue), long-dashed-short-dashed~(black) curves 
denote $V/E^2$ for
I~(Schwarzschild metric with $a=0$ and $m=1$), 
II~(regular black hole with $a=1$ and $m=1$),  
III~(one-way traversable wormhole with $a=2$ and $m=1$),
IV~(traversable wormhole with $a=2.5$ and $m=1$),
V~(traversable wormhole with a marginally unstable photon sphere with $a=3$ and $m=1$),
VI~(traversable wormhole with $a=10$ and $m=1$), and
VII~(Ellis-Bronnikov wormhole metric with $a=1$ and $m=0$), respectively.}
\end{figure}

A deflection angle $\alpha$ is obtained as, from Eq.~(\ref{eq:trajectory}),
\begin{eqnarray}\label{eq:defalp}
\alpha=I(r_0)-\pi,
\end{eqnarray}
where 
\begin{eqnarray}
I(r_0)\equiv 2 \int^\infty_{r_{0}} \frac{dr}{\sqrt{\frac{F(r)C(r)}{B(r)}}}.
\end{eqnarray}

\subsection{Standard radial coordinate~$\rho$}
By using the standard radial coordinate~$\rho \equiv \sqrt{r^2+a^2}$,
the line element~(\ref{eq:line_element}) is rewritten as
\begin{eqnarray}\label{eq:line_element_standard}
ds^2
=-A(\rho)dt^2 +\tilde{B}(\rho)d\rho^2 + C(\rho)(d\vartheta^2+\sin^2\vartheta d\varphi^2), \nonumber\\
\end{eqnarray}
where $A(\rho)$, $\tilde{B}(\rho)$, and $C(\rho)$ are given by
\begin{eqnarray}
A(\rho)&\equiv& 1- \frac{2m}{\rho}, \\
\tilde{B}(\rho)&\equiv& \frac{1}{\left(1- \frac{2m}{\rho}\right) \left( 1-\frac{a^2}{\rho^2} \right) }, \\ \label{eq:C_standard}
C(\rho)&\equiv& \rho^2.
\end{eqnarray}
The trajectory of a light ray is expressed as
\begin{eqnarray}
\dot{\rho}^2+\tilde{V}(\rho)=0,
\end{eqnarray}
where $\tilde{V}(\rho)$ is an effective potential for the radial coordinate $\rho$ defined as
\begin{eqnarray}
\tilde{V}(\rho)\equiv -\frac{L^2 F(\rho)}{\tilde{B}(\rho)C(\rho)}.
\end{eqnarray}
The deflection angle $\alpha$~(\ref{eq:defalp}) is rewritten as 
\begin{eqnarray}\label{eq:defalp_standard}
\alpha=I(\rho_0)-\pi,
\end{eqnarray}
where $I(\rho_0)$ is given by
\begin{eqnarray}
I(\rho_0)
&=&2 \int^\infty_{\rho_{0}} \frac{d\rho}{\sqrt{\frac{F(\rho)C(\rho)}{\tilde{B}(\rho)}}} \nonumber\\
&=&2 \int^\infty_{\rho_{0}} \frac{d\rho}{\sqrt{ \left( \frac{A_0 \rho^2}{\rho_0^2}-A \right) \left( \rho^2-a^2 \right) }}.
\end{eqnarray}

\section{Deflection angle in the strong deflection limit}
In this section, we investigate the deflection angle in the strong deflection limit $r_0 \rightarrow r_\mathrm{m}$ or $b \rightarrow b_\mathrm{m}$.
By using a variable $z$ defined by
\begin{eqnarray}\label{eq:z}
z\equiv \frac{g_{tt}(r)-g_{tt}(r_0)}{1-g_{tt}(r_0)}=1-\frac{\sqrt{r_0^2 +a^2}}{\sqrt{r^2 +a^2}},
\end{eqnarray}
$I(r_0)$ can be expressed by
\begin{eqnarray}
I(r_0)
=\int^1_0 \frac{2 \left( a^2+r_0^2 \right)^{3/4} dz}{\sqrt{r_0^2+2a^2 z-a^2 z^2} \sqrt{c_1(r_0) z + c_2(r_0) z^2 -2m z^3}}, \nonumber\\
\end{eqnarray}
where 
\begin{eqnarray}
&&c_1(r_0) \equiv 2 \left( \sqrt{a^2+r_0^2}-3m \right), \\
&&c_2(r_0) \equiv 6m-\sqrt{a^2+r_0^2}.
\end{eqnarray}

\subsection{Case of $a<3m$}
For $a<3m$, from $V_\mathrm{m}=V^{\prime}_\mathrm{m}=0$, 
we find a circular light orbit at $r_\mathrm{m}=\sqrt{9m^2-a^2}$ with $b_\mathrm{m}=3\sqrt{3}m$.
From $V^{\prime \prime }_\mathrm{m}<0$, the circular light orbit forms a photon sphere. 
We express $I(r_0)$ as
\begin{eqnarray}
I(r_0)
=\int^1_0 R(z,r_0) f(z,r_0) dz,
\end{eqnarray}
where $R(z,r_0)$ and $f(z,r_0)$ are defined by
\begin{eqnarray}
R(z,r_0) \equiv \frac{2 \left( a^2+r_0^2 \right)^{3/4}}{\sqrt{r_0^2+2a^2 z-a^2 z^2} }
\end{eqnarray}
and 
\begin{eqnarray}
f(z,r_0) 
\equiv \frac{1}{\sqrt{c_1(r_0) z + c_2(r_0) z^2 -2m z^3}},
\end{eqnarray}
respectively.
$c_1(r_0)$ and $c_2(r_0)$ are expanded in power of $r_0-r_\mathrm{m}$ as
\begin{eqnarray}\label{eq:cm1}
&&c_1(r_0) = \frac{2 \sqrt{9 m^2-a^2}}{3 m} \left( r_0-r_\mathrm{m} \right) + O((r_0-r_\mathrm{m})^2), \qquad \\ \label{eq:cm2}
&&c_2(r_0) = 3m +O(r_0-r_\mathrm{m}).
\end{eqnarray}
Thus, $f(z,r_0)$ diverges as $z^{-1}$ in the strong deflection limit $r_0 \rightarrow r_\mathrm{m}$.  

We separate $I(r_0)$ as 
\begin{eqnarray}
I(r_0)=I_\mathrm{D}(r_0)+I_\mathrm{R}(r_0),
\end{eqnarray}
where $I_\mathrm{D}(r_0)$ is a divergent term defined by
\begin{eqnarray}
I_\mathrm{D}(r_0)
\equiv \int^1_0 R(0,r_\mathrm{m}) f_\mathrm{D}(z,r_0) dz,
\end{eqnarray}
where $f_\mathrm{D}(z,r_0)$ is defined as 
\begin{eqnarray}
f_\mathrm{D}(z,r_0) \equiv \frac{1}{\sqrt{c_1(r_0) z + c_2(r_0) z^2 }}
\end{eqnarray}
and $R(0,r_\mathrm{m})$ is given by
\begin{eqnarray}
R(0,r_\mathrm{m}) =\frac{6m\sqrt{3m}}{\sqrt{9m^2-a^2}}.
\end{eqnarray}
The divergent term $I_\mathrm{D}(r_0)$ yields 
\begin{eqnarray}
I_\mathrm{D}(r_0)
=\frac{2R(0,r_\mathrm{m})}{\sqrt{c_2(r_0)}} \log \frac{\sqrt{c_2(r_0)}+\sqrt{c_1(r_0)+c_2(r_0)}}{\sqrt{c_1(r_0)}}. \nonumber\\
\end{eqnarray}
From Eqs.~(\ref{eq:cm1}), (\ref{eq:cm2}) and 
\begin{eqnarray}
b(r_0)=b_\mathrm{m}+\frac{9m^2-a^2}{6 \sqrt{3}m^3}(r_0-r_\mathrm{m})^2+O((r_0-r_\mathrm{m})^3), \nonumber\\
\end{eqnarray}
the divergent term $I_\mathrm{D}(r_0)$ becomes in the strong deflection limit $b \rightarrow b_\mathrm{m}$
\begin{eqnarray}
I_\mathrm{D}(r_0)
&=&-\frac{3m}{\sqrt{9m^2-a^2}} \log \left( \frac{b}{b_\mathrm{m}}-1 \right) + \frac{3m}{\sqrt{9m^2-a^2}} \log 6  \nonumber\\
&&+O \left( \left( \frac{b}{b_\mathrm{m}}-1 \right) \log \left( \frac{b}{b_\mathrm{m}}-1 \right) \right).
\end{eqnarray}

We define a regular part $I_\mathrm{R}(r_0)$ as
\begin{eqnarray}
I_\mathrm{R}(r_0) \equiv \int^1_0  g(z,r_0) dz,
\end{eqnarray}
where $g(z,r_0)$ defined by
\begin{equation}
g(z,r_0)\equiv R(z,r_0) f(z,r_0) -R(0,r_\mathrm{m}) f_\mathrm{D}(z,r_0)
\end{equation}
and it is expanded in the power of $r_0-r_\mathrm{m}$ as
\begin{eqnarray}
I_\mathrm{R}(r_0) = \sum^\infty_{j=0} \frac{1}{j!} (r_0-r_\mathrm{m})^j \int^1_0 \left. \frac{\partial^j g}{\partial r_0^j} \right|_{r_0=r_\mathrm{m}} dz.
\end{eqnarray}
We are interested in the term of $j=0$ given by
\begin{eqnarray}\label{eq:I_R0}
I_\mathrm{R}(r_0) = \int^1_0 g(z,r_\mathrm{m}) dz,
\end{eqnarray}
where $g(z,r_\mathrm{m})$ is expressed by
\begin{eqnarray}\label{eq:g}
g(z,r_\mathrm{m})
&=&\left[ \frac{\sqrt{3}}{\sqrt{9m^2-a^2+a^2 (2-z) z} \sqrt{3-2z}} \right. \nonumber\\
&&\left. - \frac{1}{\sqrt{9m^2-a^2}} \right] \frac{6m}{z}.
\end{eqnarray}

Therefore, the deflection angle in the strong deflection limit has the form of Eq.~(\ref{eq:def0}) 
and the parameters~$\bar{a}$ and $\bar{b}$ are obtained as
\begin{eqnarray}\label{eq:bar_a}
&&\bar{a}=\frac{3m}{\sqrt{9m^2-a^2}}, \\ \label{eq:bar_b}
&&\bar{b}=\frac{3m}{\sqrt{9m^2-a^2}} \log 6 +I_\mathrm{R} -\pi.
\end{eqnarray}
Figure 2 shows $\bar{a}$ and $\bar{b}$ as a function of $a/m$.
Note that $I_\mathrm{R}$ is obtained numerically in usual. 
In the Schwarzschild case of $a=0$, we obtain $I_\mathrm{R}=\log \left[ 36(7-4\sqrt{3}) \right]$, 
$\bar{a}=1$, and $\bar{b}=\log \left[ 216(7-4\sqrt{3}) \right] -\pi \sim -0.40023$ analytically.
They are equivalent to a result in Refs.~\cite{Darwin_1959,Bozza_Capozziello_Iovane_Scarpetta_2001,Bozza:2002zj}. 
\begin{figure}[htbp]
\begin{center}
\includegraphics[width=87mm]{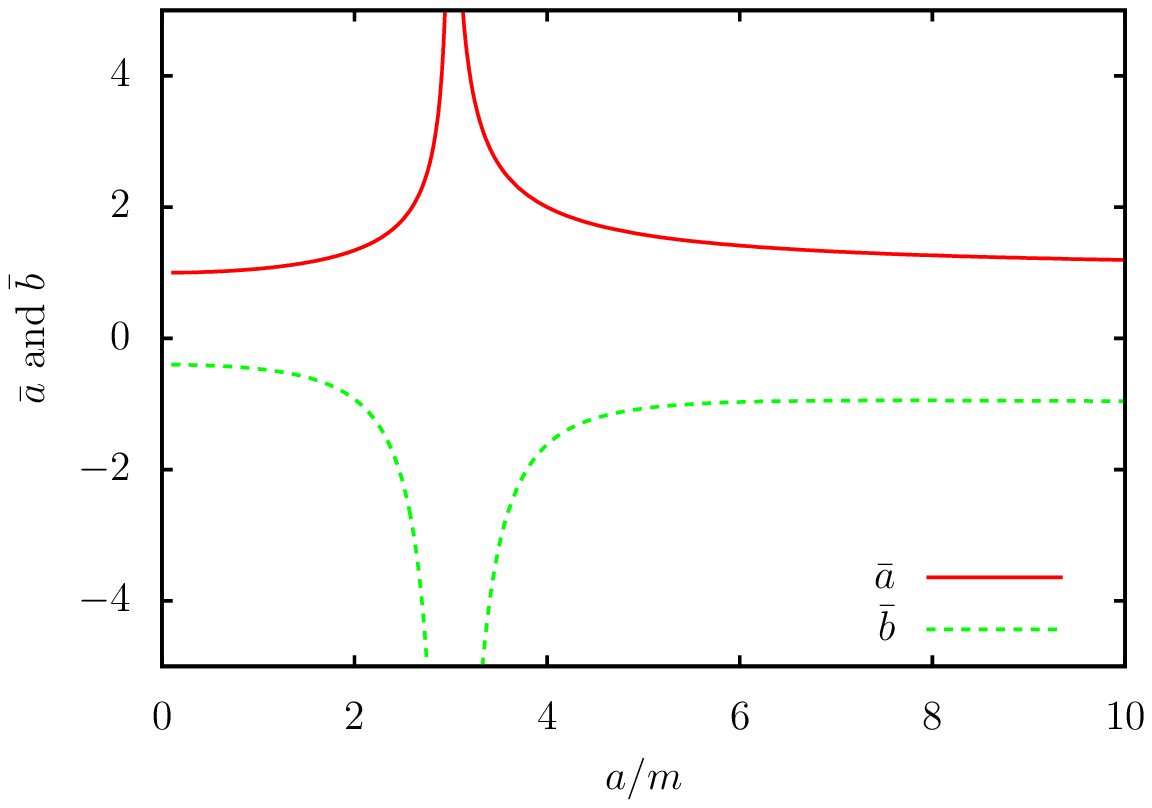}
\end{center}
\caption{
The solid~(red) and broken~(green) curves denote $\bar{a}$ and $\bar{b}$, respectively, as a function of $a/m$.}
\end{figure}

We notice that $\bar{a}$ given by Eq.~(\ref{eq:bar_a}) is the same as Eq.~(27) shown by Nascimento~\textit{et al.}~\cite{Nascimento:2020ime}.
Nascimento~\textit{et al.} chose a variable~$z_{*}\equiv 1-r_0/r$ as shown Eq.~(17)
and the variable $z_{*}$ gave a complicated form of $\bar{b}$ which is calculated by Eqs.~(28) and (31) in Ref.~\cite{Nascimento:2020ime}.
We have chosen the variable $z$ instead of $z_{*}$ to get the simpler form of $\bar{b}$.
We have confirmed our result by comparing Eq.~(1.1) with Eq.~(2.13) in numerical as shown Fig.~3. 
\begin{figure}[htbp]
\begin{center}
\includegraphics[width=87mm]{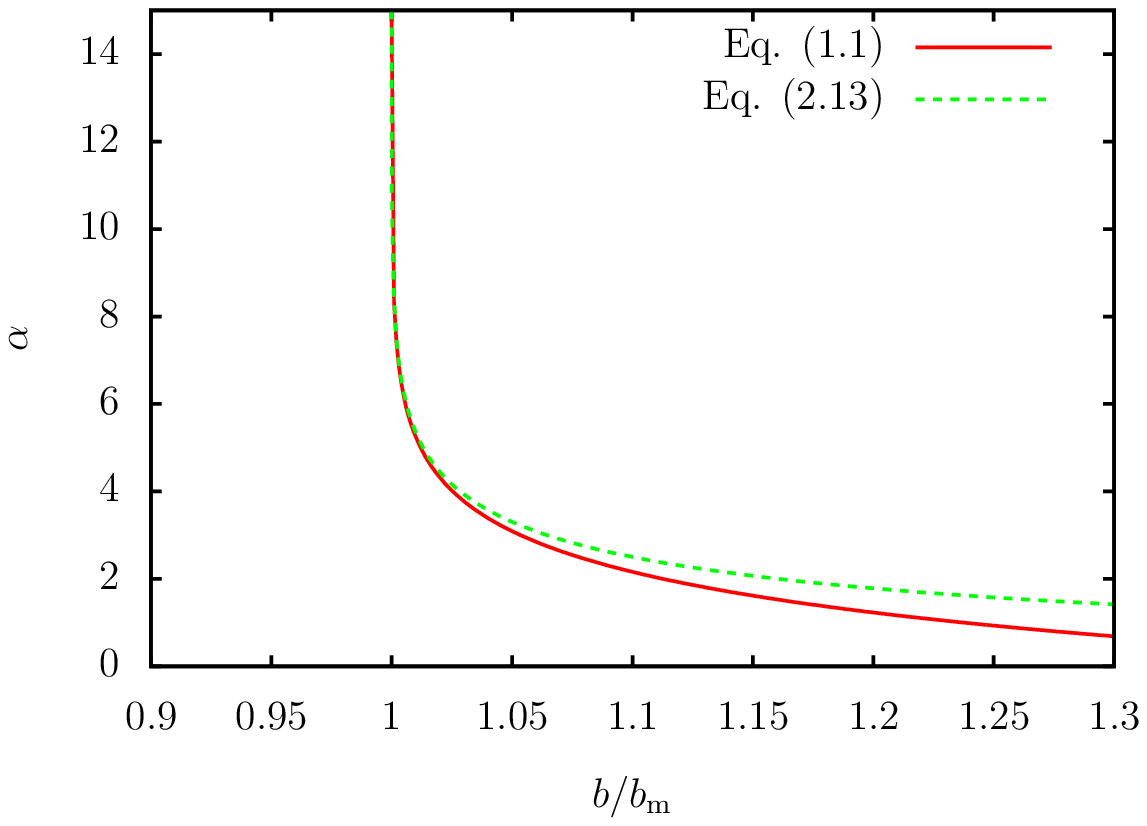}
\end{center}
\caption{
The deflection angle in the case of $a=2m$. A solid (red) curve is calculated by the strong deflection limit~(1.1) 
and a dashed (green) curve is calculated by Eq.~(2.13) in numerical.}
\end{figure}

\subsection{Case of $a>3m$}
In the case of $a>3m$,
the wormhole throat at $r=0$
correspond with the photon sphere with $V^{\prime \prime }_\mathrm{m}<0$.
Notice $r_\mathrm{m}=0$ and
\begin{eqnarray}
b_\mathrm{m}=\sqrt{\frac{a^3}{a-2m}}
\end{eqnarray}
due to $V_\mathrm{m}=V^{\prime}_\mathrm{m}=0$.
We express $I(r_0)$ as
\begin{eqnarray}
I(r_0)
=\int^1_0 S(r_0) h(z,r_0) dz,
\end{eqnarray}
where $S(r_0)$ and $h(z,r_0)$ are defined by
\begin{eqnarray}
S(r_0) \equiv 2 \left( a^2+r_0^2 \right)^{3/4}
\end{eqnarray}
and
\begin{eqnarray}
&&h(z,r_0) \nonumber\\
&&\equiv \frac{1}{\sqrt{c_3(r_0)z +c_4(r_0)z^2 +c_5(r_0)z^3 +c_6(r_0)z^4 +2ma^2 z^5}}, \nonumber\\
\end{eqnarray}
respectively, and where $c_3(r_0)$, $c_4(r_0)$, $c_5(r_0)$, and $c_6(r_0)$ are defined by 
\begin{eqnarray}
&&c_3(r_0) \equiv r_0^2 c_1,  \\
&&c_4(r_0) \equiv r_0^2 c_2 +2a^2 c_1, \\
&&c_5(r_0) \equiv -2 r_0^2 m +2a^2c_2-a^2c_1, \\
&&c_6(r_0) \equiv -4a^2m-a^2c_2.
\end{eqnarray}
We can expand $c_3(r_0)$ and $c_4(r_0)$ in the power of $r_0-r_\mathrm{m}$ as
\begin{eqnarray}\label{eq:c3m}
&&c_3(r_0) = 2(a-3m)(r_0-r_\mathrm{m})^2 +O((r_0-r_\mathrm{m})^3), \quad \\ \label{eq:c4m}
&&c_4(r_0) = 4 a^2 (a-3 m) +O(r_0-r_\mathrm{m}).
\end{eqnarray}
Thus, $h(z,r_0)$ diverges as $z^{-1}$ in the strong deflection limit $r_0 \rightarrow r_\mathrm{m}$.  


In the case, we separate $I(r_0)$ as
\begin{eqnarray}
I(r_0)=I_\mathrm{d}(r_0)+I_\mathrm{r}(r_0),
\end{eqnarray}
where $I_\mathrm{d}(r_0)$ is a divergent term and $I_\mathrm{r}(r_0)$ is a regular part.
We define the divergent term $I_\mathrm{d}(r_0)$ as
\begin{eqnarray}
I_\mathrm{d}(r_0)
\equiv \int^1_0 S(r_\mathrm{m}) h_\mathrm{d}(z,r_0) dz,
\end{eqnarray}
where $h_\mathrm{d}(z,r_0)$ is defined by
\begin{eqnarray}\label{eq:hD1}
h_\mathrm{d}(z,r_0) \equiv \frac{1}{\sqrt{c_3(r_0) z + c_4(r_0) z^2 }}
\end{eqnarray}
and $S(r_\mathrm{m})$ is 
\begin{eqnarray}
S(r_\mathrm{m})=2a^{3/2}.
\end{eqnarray}
The divergent term is obtained as
\begin{eqnarray}
I_\mathrm{d}(r_0)
=\frac{2S(r_\mathrm{m})}{\sqrt{c_4(r_0)}} \log \frac{\sqrt{c_4(r_0)}+\sqrt{c_3(r_0)+c_4(r_0)}}{\sqrt{c_3(r_0)}}. \nonumber\\
\end{eqnarray}
From 
\begin{eqnarray}
b(r_0)=b_\mathrm{m}+\frac{a-3m}{2 \sqrt{a}(a-2m)^{3/2}}(r_0-r_\mathrm{m})^2+O((r_0-r_\mathrm{m})^3) \nonumber\\
\end{eqnarray}
and Eqs.~(\ref{eq:c3m}) and (\ref{eq:c4m}),
we get the divergent term in the strong deflection limit~$b \rightarrow b_\mathrm{m}$ as
\begin{eqnarray}
I_\mathrm{d}(r_0)
&=&-\sqrt{\frac{a}{a-3m}} \log \left( \frac{b}{b_\mathrm{m}}-1 \right) \nonumber\\
&&+ \sqrt{\frac{a}{a-3m}} \log \frac{4(a-3m)}{a-2m}  \nonumber\\
&&+O \left( \left( \frac{b}{b_\mathrm{m}}-1 \right) \log \left( \frac{b}{b_\mathrm{m}}-1 \right) \right).
\end{eqnarray}

We define the regular term $I_\mathrm{r}(r_0)$ as
\begin{eqnarray}
I_\mathrm{r}(r_0) \equiv \int^1_0  k(z,r_0) dz,
\end{eqnarray}
where $k(z,r_0)$ is given by
\begin{eqnarray}
k(z,r_0)\equiv S(r_0) h(z,r_0) -S(r_\mathrm{m}) h_\mathrm{d}(z,r_0).
\end{eqnarray}
We expand $I_\mathrm{r}(r_0)$ in the power of $r_0-r_\mathrm{m}$ as
\begin{eqnarray}
I_\mathrm{r}(r_0) = \sum^\infty_{j=0} \frac{1}{j!} (r_0-r_\mathrm{m})^j \int^1_0 \left. \frac{\partial^j k}{\partial r_0^j} \right|_{r_0=r_\mathrm{m}} dz,
\end{eqnarray}
and the term of $j=0$ is given by
\begin{eqnarray}\label{eq:I_R}
I_\mathrm{r} = \int^1_0 k(z,r_\mathrm{m}) dz,
\end{eqnarray}
where $k(z,r_\mathrm{m})$ is 
\begin{eqnarray}
k(z,r_\mathrm{m})
&=&\left[ \frac{2}{\sqrt{2-z}\sqrt{2 a-6 m +(6 m-a)z-2 m z^2} } \right. \nonumber\\
&&\left. - \frac{1}{\sqrt{a-3m}} \right] \frac{\sqrt{a}}{z}.
\end{eqnarray}

The deflection angle in the strong deflection limit has the form of Eq.~(\ref{eq:def0}) and 
parameters $\bar{a}$ and $\bar{b}$ are obtained as
\begin{eqnarray}
&&\bar{a}=\sqrt{\frac{a}{a-3m}}, \\
&&\bar{b}=\sqrt{\frac{a}{a-3m}} \log \frac{4(a-3m)}{a-2m} +I_\mathrm{r} -\pi
\end{eqnarray}
and they have been shown in Fig.~2.
We have confirmed our result in numerical as shown Fig.~4.
When $a\neq 0$ and $m=0$, the metric coincides with the Ellis-Bronnikov wormhole metric which 
is a solution of Einstein and scalar field equations~\cite{Ellis_1973,Bronnikov_1973,Martinez:2020hjm}.
Gravitational lensing by the Ellis-Bronnikov wormhole has investigated eagerly~\cite{Chetouani_Clement_1984,Perlick:2003vg,Nandi:2006ds,Muller:2008zza,Dey_Sen_2008,Abe_2010,Bhattacharya:2010zzb,
Toki_Kitamura_Asada_Abe_2011,Gibbons_Vyska_2012,Nakajima_Asada_2012,Tsukamoto_Harada_Yajima_2012,Tsukamoto_Harada_2013,Yoo_Harada_Tsukamoto_2013,Takahashi_Asada_2013,Izumi_2013,Nakajima:2014nba,Perlick:2014zwa,Bozza:2015haa,
Bozza:2015wbw,Lukmanova_2016,Tsukamoto:2016qro,Tsukamoto:2016zdu,Nandi:2016uzg,Jusufi:2017gyu,Bozza:2017dkv,Tsukamoto:2017hva,Tsukamoto:2017edq,Bhattacharya:2019kkb,Shaikh:2019jfr,Bozza:2020ubm}.
See Ref.~\cite{Tsukamoto:2016qro} and references therein for the details of the Ellis-Bronnikov wormhole.
In the case, we obtain $I_\mathrm{r}=\log2$, 
and then $\bar{a}=1$ and $\bar{b}=3\log 2 -\pi \sim -1.06215$.
This is equivalent to a result in Refs.~\cite{Tsukamoto:2016qro,Tsukamoto:2016jzh,Bhattacharya:2019kkb}.
\begin{figure}[htbp]
\begin{center}
\includegraphics[width=87mm]{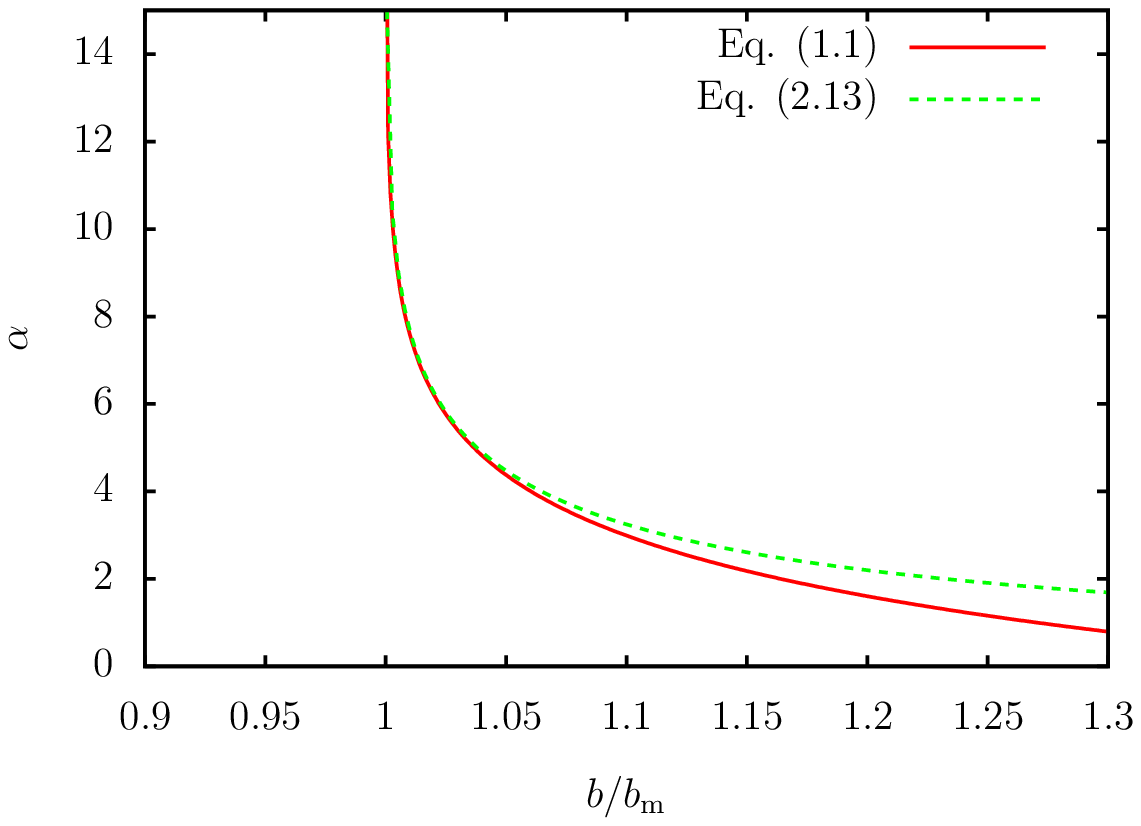}
\end{center}
\caption{
The deflection angle for $a=4m$. A solid (red) curve is calculated by the strong deflection limit~(1.1) 
and a dashed (green) curve is calculated by Eq.~(2.13) in numerical.}
\end{figure}

\subsection{Case of $a=3m$}
In the case of $a=3m$, the photon sphere at $r=r_\mathrm{m}=0$, which is correspond with a wormhole throat, is marginally unstable
since the light ray with $b_\mathrm{m}=3\sqrt{3}m$ satisfies $V_\mathrm{m}=V^{\prime}_\mathrm{m}=V^{\prime \prime }_\mathrm{m}=V^{\prime \prime \prime }_\mathrm{m}=0$ and $V^{\prime \prime \prime \prime }_\mathrm{m}<0$. 
From $c_3(r_0)$, $c_4(r_0)$, $c_5(r_0)$, and $c_6(r_0)$ which are expanded in the power of $r_0-r_\mathrm{m}$ as
\begin{eqnarray}
&&c_3(r_0) = \frac{1}{3m}(r_0-r_\mathrm{m})^4 +O((r_0-r_\mathrm{m})^5), \quad \\
&&c_4(r_0) = 9m (r_0-r_\mathrm{m})^2 +O((r_0-r_\mathrm{m})^3),\\
&&c_5(r_0) = 54m^3 +O(r_0-r_\mathrm{m}),\\
&&c_6(r_0) = -63m^3 +O(r_0-r_\mathrm{m}),
\end{eqnarray}
$h(z,r_\mathrm{m})$ is obtained as
\begin{eqnarray}
h(z,r_\mathrm{m}) = \frac{1}{3m^{3/2}z^{3/2}\sqrt{6-7z+2z^2}}.
\end{eqnarray}
Thus, in the case, we define $h_\mathrm{d}$ as not Eq.~(\ref{eq:hD1}) but 
\begin{eqnarray}
h_\mathrm{d}(z)\equiv \frac{\sqrt{6}}{18m^{3/2}z^{3/2}}.
\end{eqnarray}
The divergent term $I_\mathrm{d}$ is given by  
\begin{eqnarray}
I_\mathrm{d}
&\sim& \left. \frac{2\sqrt{2}}{\sqrt{z}} \right|_{z=0} -2\sqrt{2} \nonumber\\
&\sim& \left. \frac{2\sqrt{2}}{\sqrt{\frac{\rho_0}{\rho_\mathrm{m}}-1}} \right|_{\rho_0=\rho_\mathrm{m}} -2\sqrt{2}. 
\end{eqnarray}
Notice that the position of the marginally unstable photon sphere or the wormhole throat is given by $\rho=\rho_\mathrm{m}=3m$ 
and the variable $z$ is expressed by
\begin{eqnarray}\label{eq:z2}
z=1-\frac{\rho_0}{\rho}
\end{eqnarray}
by using the standard radial coordinate $\rho$.
From the relation between the impact parameter $b$ and the closest distance $\rho_0$
\begin{eqnarray}
b-b_\mathrm{m}
=
\frac{\sqrt{3}}{2 m}(\rho_0-\rho_\mathrm{m})^2+O((\rho_0-\rho_\mathrm{m})^3),
\end{eqnarray}
the divergent term $I_\mathrm{d}$ is given by
\begin{eqnarray}
I_\mathrm{d}
\sim \frac{2^{5/4}3^{1/4}}{\left(\frac{b}{b_\mathrm{m}}-1\right)^{1/4}} -2\sqrt{2}. 
\end{eqnarray}


From 
\begin{eqnarray}
k(z,r_\mathrm{m})
=\frac{2\sqrt{3}}{z^{3/2}} \left( \frac{1}{\sqrt{6-7z+2z^2}}-\frac{\sqrt{6}}{6} \right),
\end{eqnarray}
the regular term~(\ref{eq:I_R}) is obtained as
\begin{eqnarray}
I_\mathrm{r}=2\sqrt{2}\left(1+ K\left( \sqrt{\frac{1}{6}} \right) -E\left( \sqrt{\frac{1}{6}} \right) \right), \nonumber\\
\end{eqnarray}
where $K(x)$ and $E(x)$ are complete elliptic integrals of the first and second kinds defined by  
\begin{eqnarray}
K(x)\equiv \int^\frac{\pi}{2}_0 \frac{d\theta}{\sqrt{1-x^2 \sin^2 \theta}}
\end{eqnarray}
and
\begin{eqnarray}
E(x)\equiv \int^\frac{\pi}{2}_0 \sqrt{1-x^2 \sin^2 \theta}d\theta,
\end{eqnarray}
respectively.

The parameters $\bar{c}$ and $\bar{d}$ in the deflection angle in the strong deflection limit 
with the form of Eq.~(\ref{eq:def02}) are obtained as
\begin{eqnarray}
\bar{c}&=&2^{5/4}3^{1/4} \sim 3.13017, \\
\bar{d}&=& 2\sqrt{2} \left( K\left( \sqrt{\frac{1}{6}} \right) -E\left( \sqrt{\frac{1}{6}} \right) \right)- \pi \nonumber\\
&\sim& -2.74546. \qquad 
\end{eqnarray}
We confirm the deflection angle in the strong deflection limit by comparing Eq.~(1.2) with Eq.~(2.13).
The error of the deflection angles for $\alpha \gtrsim \pi$ is small enough as shown Fig.~5.
\begin{figure}[htbp]
\begin{center}
\includegraphics[width=87mm]{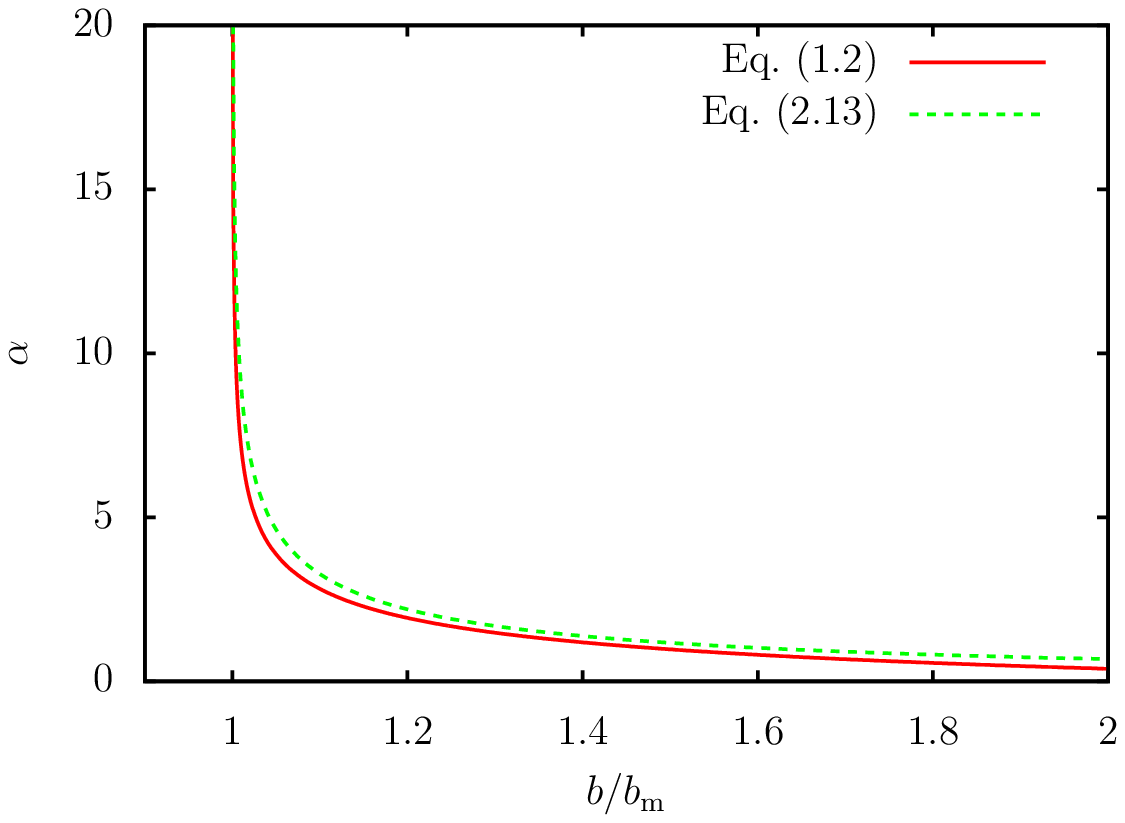}
\end{center}
\caption{
The deflection angle for $a=3m$. A solid (red) curve is calculated by the strong deflection limit~(1.2) and a dashed (green) curve is calculated by Eq.~(2.13) in numerical.}
\end{figure}

\section{Observables in the strong deflection limit}
We consider that a light ray, which is emitted by a source~S with a source angle~$\phi$, 
is scattered by a lens~L with an deflection angle~$\alpha$, 
and that it is observed as an image~$I$ with an image angle~$\theta$ by an observer~O. 
The lens configuration is shown in Fig.~6.   
\begin{figure}[htbp]
\begin{center}
\includegraphics[width=87mm]{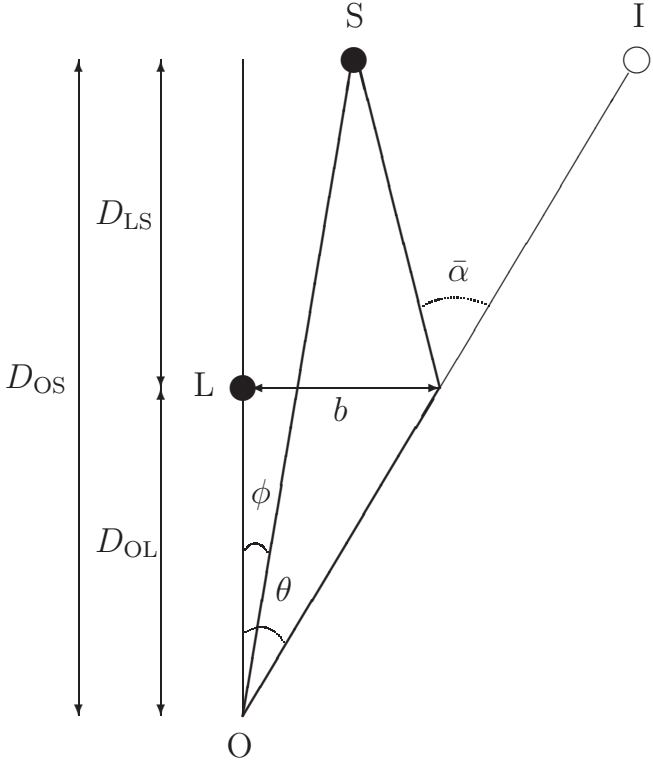}
\end{center}
\caption{
A lens configuration. A light ray emitted by a source~S at a source angle~$\phi$
is scattered by a lens~L with an effective deflection angle~$\bar{\alpha}$, 
and it is observed as an image~$I$ with an image angle~$\theta$ by an observer~O.
$b$ is the impact parameter of the light ray. 
$D_{\mathrm{OS}}$, $D_{\mathrm{LS}}$, and $D_{\mathrm{OL}}=D_{\mathrm{OS}}-D_{\mathrm{LS}}$ are distances between the observer and the source, between the lens and the source, and between the observer and the lens, respectively. 
}
\end{figure}
We define an effective deflection angle as
\begin{eqnarray}
\bar{\alpha}= \alpha \:\:\:  \mathrm{mod} \:\: 2\pi
\end{eqnarray}
and we assume $\phi \ll 1$, $\bar{\alpha} \ll 1$, and $\theta=b/D_{\mathrm{OL}} \ll 1$,
where $D_{\mathrm{OL}}$ is a distance between the observer and the lens. 
A small angle lens equation~\cite{Bozza:2008ev} is obtained as
\begin{eqnarray}\label{eq:Lens}
D_{\mathrm{LS}} \bar{\alpha}= D_{\mathrm{OS}} \left( \theta-\phi \right),
\end{eqnarray}
where $D_{\mathrm{LS}}$ and $D_{\mathrm{OS}}=D_{\mathrm{OL}}+D_{\mathrm{LS}}$ are distances between the lens and the source and between the observer and the source, respectively.
By using a winding number~$n$ which is a non-negative integer, 
the deflection angle is expressed by
\begin{eqnarray}\label{eq:alpha02}
\alpha=\bar{\alpha}+2\pi n.
\end{eqnarray}
We expand the deflection angle $\alpha(\theta)$ around $\theta=\theta^0_n$ as  
\begin{eqnarray}
\alpha(\theta)&=&\alpha \left(\theta_n^0 \right) + \left. \frac{d \alpha}{d \theta} \right|_{\theta=\theta_n^0}  \left(\theta-\theta^0_n \right) \nonumber\\
&&+O\left( \left(\theta-\theta^0_n \right)^2 \right),
\end{eqnarray} 
where $\theta_n^0$ is defined by
\begin{eqnarray}\label{eq:2pi_n}
\alpha(\theta_n^0) = 2\pi n. 
\end{eqnarray}

\subsection{$a\neq 3m$}
In the case of $a\neq 3m$, the deflection angle in the strong deflection limit is written in
\begin{eqnarray}\label{eq:alpha}
\alpha(\theta) 
&=& -\bar{a} \log \left( \frac{\theta}{\theta_\infty}-1 \right) +\bar{b} \nonumber\\
&&+O\left( \left( \frac{\theta}{\theta_\infty}-1 \right) \log \left( \frac{\theta}{\theta_\infty}-1 \right) \right),
\end{eqnarray}
where $\theta_\infty$ is the image angle of the photon sphere defined by $\theta_\infty \equiv b_\mathrm{m}/D_{\mathrm{OL}}$
and we obtain 
\begin{eqnarray}\label{eq:dalpha}
\left. \frac{d\alpha}{d\theta}\right|_{\theta=\theta^0_n}=-\frac{\bar{a}}{\theta^0_n-\theta_{\infty}}.
\end{eqnarray}
From Eqs.~(\ref{eq:2pi_n}) and (\ref{eq:alpha}), we get
\begin{eqnarray}\label{eq:theta0}
\theta^0_n= \left( 1+ e^{\frac{\bar{b}-2\pi n}{\bar{a}}} \right) \theta_\infty.
\end{eqnarray}
The effective deflection angle~$\bar{\alpha}$ is given by, from Eqs.~(\ref{eq:alpha02})-(\ref{eq:2pi_n}), (\ref{eq:dalpha}), and (\ref{eq:theta0}),
\begin{eqnarray}\label{eq:eff_alpha}
\bar{\alpha}(\theta_n)=\frac{\bar{a}}{\theta_\infty e^{\frac{\bar{b}-2\pi n}{\bar{a}}}} \left(  \theta^0_n-\theta_n \right),
\end{eqnarray}
where $\theta=\theta_n$ is a solution of the lens equation~(\ref{eq:Lens}) with the winding number $n$.

By substituting the effective deflection angle~(\ref{eq:eff_alpha}) into the lens equation~(\ref{eq:Lens}),  
the image angle is obtained as
\begin{eqnarray}\label{eq:theta_n2}
\theta_n(\phi)\sim \theta^0_n+\frac{\theta_\infty e^{\frac{\bar{b}-2\pi n}{\bar{a}}} D_{\mathrm{OS}} \left( \phi-\theta^0_n \right)}{\bar{a}D_{\mathrm{LS}}}.
\end{eqnarray} 
When the observer, the lens, and the source are aligned in a line, ring images called relativistic Einstein rings are formed. The ring angle $\theta_{\mathrm{E} n}$ is 
\begin{eqnarray}
\theta_{\mathrm{E} n} \equiv \theta_n(0)= \left( 1 -\frac{\theta_\infty e^{\frac{\bar{b}-2\pi n}{\bar{a}}} D_{\mathrm{OS}} }{\bar{a}D_{\mathrm{LS}}} \right) \theta^0_n.
\end{eqnarray} 
The difference of the image angles between the outermost relativistic images and the photon sphere is obtained as
\begin{eqnarray}
\mathrm{s}\equiv \theta_1-\theta_\infty \sim \theta^0_1- \theta^0_\infty= \theta_\infty e^{\frac{\bar{b}-2\pi}{\bar{a}}}.
\end{eqnarray}

The magnification of the image is obtained as
\begin{eqnarray}
\mu_n 
&\equiv& \frac{\theta_n}{\phi} \frac{d\theta_n}{d\phi} \nonumber\\
&\sim&  \frac{\theta_\infty^2 D_{\mathrm{OS}} \left( 1+ e^{\frac{\bar{b}-2\pi n}{\bar{a}}} \right) e^{\frac{\bar{b}-2\pi n}{\bar{a}}}}{\phi \bar{a} D_{\mathrm{LS}} }.
\end{eqnarray}
The sum of the magnifications of the images from $n=1$ to $\infty$ is given by
\begin{eqnarray}
\sum^\infty_{n=1} \mu_n  
\sim  \frac{\theta_\infty^2 D_{\mathrm{OS}} \left( 1+ e^{\frac{2\pi}{\bar{a}}}+e^{\frac{\bar{b}}{\bar{a}}} \right) e^{\frac{\bar{b}}{\bar{a}}}}
{\phi \bar{a} D_{\mathrm{LS}} \left( e^{\frac{4\pi}{\bar{a}}} -1 \right) }.
\end{eqnarray}
The ratio of the magnification of the outermost relativistic image to the others 
\begin{eqnarray}
\mathrm{r}\equiv \frac{\mu_1}{\sum^\infty_{n=2} \mu_n}\sim \frac{\left( e^{\frac{4\pi}{\bar{a}}}-1 \right) \left( e^{\frac{2\pi}{\bar{a}}}+e^{\frac{\bar{b}}{\bar{a}}} \right)}
{ e^{\frac{4\pi}{\bar{a}}}+ e^{\frac{2\pi}{\bar{a}}}+e^{\frac{\bar{b}}{\bar{a}}}},
\end{eqnarray}
where the magnification without the outermost relativistic image is given by
\begin{eqnarray}
\sum^\infty_{n=2} \mu_n  
\sim  \frac{\theta_\infty^2 D_{\mathrm{OS}} \left( e^{\frac{4\pi}{\bar{a}}}+ e^{\frac{2\pi}{\bar{a}}}+e^{\frac{\bar{b}}{\bar{a}}} \right) e^{\frac{\bar{b}-4\pi}{\bar{a}}} }
{\phi \bar{a} D_{\mathrm{LS}} \left( e^{\frac{4\pi}{\bar{a}}} -1 \right) }.
\end{eqnarray}

\subsection{$a=3m$}
In the case of $a=3m$, the deflection angle in the strong deflection limit is given by
\begin{eqnarray}\label{eq:alpha2}
\alpha(\theta)= \frac{\bar{c}}{\left( \frac{\theta}{\theta_\infty}-1 \right)^{1/4}} +\bar{d}
\end{eqnarray}
and it yields  
\begin{eqnarray}\label{eq:dalpha2}
\left. \frac{d\alpha}{d\theta}\right|_{\theta=\theta^0_n}=-\frac{\bar{c}}{4\theta_\infty} \left( \frac{\theta^0_n}{\theta_\infty}-1 \right)^{-\frac{5}{4}}.
\end{eqnarray}
We obtain, from Eqs.~(\ref{eq:2pi_n}) and (\ref{eq:alpha2}), 
\begin{eqnarray}\label{eq:theta0n2}
\theta^0_n= \left[ 1+ \left( \frac{\bar{c}}{2\pi n -\bar{d}} \right)^4 \right] \theta_\infty.
\end{eqnarray}
The effective deflection angle is obtained as, from Eqs.~(\ref{eq:alpha02})-(\ref{eq:2pi_n}), (\ref{eq:dalpha2}), and (\ref{eq:theta0n2}),
\begin{eqnarray}\label{eq:eff_alpha2}
\bar{\alpha}(\theta_n)= \frac{\left(2 \pi n -\bar{d}\right)^5}{4 \theta_\infty \bar{c}^4} \left( \theta^0_n - \theta_n \right).
\end{eqnarray}

By substituting the effective deflection angle~(\ref{eq:eff_alpha2}) into the lens equation~(\ref{eq:Lens}), the image angle and the relativistic Einstein ring angle are given by
\begin{eqnarray}
\theta_n(\phi)\sim \theta^0_n+\frac{4\bar{c}^4D_{\mathrm{OS}}\theta_\infty(\phi-\theta^0_n)}{(2\pi n-\bar{d})^5D_{\mathrm{LS}}}
\end{eqnarray}
and   
\begin{eqnarray}
\theta_{\mathrm{E}n} = \left[ 1 -\frac{4\bar{c}^4D_{\mathrm{OS}}\theta_\infty}{(2\pi n-\bar{d})^5D_{\mathrm{LS}}} \right] \theta^0_n,
\end{eqnarray}
respectively.
The difference of the image angles between the outermost relativistic image and the photon sphere is 
\begin{eqnarray}
\mathrm{s}= \theta_1-\theta_\infty = \left( \frac{\bar{c}}{2\pi -\bar{d}} \right)^4 \theta_\infty.
\end{eqnarray}

The magnification of the image is given by
\begin{eqnarray}
\mu_n \sim \frac{4 \bar{c}^4 D_{\mathrm{OS}}\theta_\infty^2F_n}{\phi D_{\mathrm{LS}}},
\end{eqnarray}
where $F_n$ is defined as
\begin{eqnarray}
F_n\equiv \frac{1+ \left( \frac{\bar{c}}{2\pi n -\bar{d}} \right)^4 }{(2\pi n-\bar{d})^5}.
\end{eqnarray}
The sum of magnifications of images from $n=1$ and $\infty$ is obtain as
\begin{eqnarray}
\sum^\infty_{n=1} \mu_n  \sim \frac{4 \bar{c}^4 D_{\mathrm{OS}}\theta_\infty^2}{\phi D_{\mathrm{LS}}} \sum^\infty_{n=1}F_n,
\end{eqnarray}
where 
\begin{eqnarray}
\sum^\infty_{n=1} F_n  \sim 1.84131 \times 10^{-5}.
\end{eqnarray}
The ratio of the magnifications of the outermost relativistic image to the other images is given by
\begin{eqnarray}
\mathrm{r}= \frac{\mu_1}{\sum^\infty_{n=2} \mu_n}\sim \frac{F_1}{\sum^\infty_{n=2} F_n}=11.2412,
\end{eqnarray}
where 
\begin{eqnarray}
F_1 \sim 1.69089 \times 10^{-5}
\end{eqnarray}
and
\begin{eqnarray}
\sum^\infty_{n=2} F_n \sim 1.50420 \times 10^{-6}.
\end{eqnarray}

\section{Gravitational lens under a weak-field approximation}
Let us review gravitational lensing under a weak-field approximation $m/\rho_0 \ll 1$ and $a/\rho_0 \ll 1$ 
in the standard radial coordinate $\rho$ briefly.
In this section, we consider not only the positive impact parameter $b$ but also negative one.
Under the weak-field approximation,
the line element given by Eqs.~(\ref{eq:line_element_standard})-(\ref{eq:C_standard}) becomes
\begin{eqnarray}\label{eq:line_element0} 
ds^2
&=&-\left( 1-\frac{2m}{\rho} \right)dt^2 + \left( 1+\frac{2m}{\rho} \right) \left( 1+\frac{a^2 }{\rho^2} \right) d\rho^2 \nonumber\\
&&+\rho^2 \left( d \vartheta^2 +\sin^2\vartheta d \varphi^2 \right)
\end{eqnarray}
and the deflection angle~$\alpha$~(\ref{eq:defalp_standard}) 
is obtained as \cite{Nascimento:2020ime,Ovgun:2020yuv}, 
\begin{eqnarray}\label{eq:alpha_weak1} 
\alpha \sim \frac{4m}{b} \qquad  \textrm{for} \quad   m \neq 0 
\end{eqnarray}
and 
\begin{eqnarray}\label{eq:alpha_weak2} 
\alpha \sim \pm \frac{\pi a^2}{4b^2}  \qquad  \textrm{for} \quad m=0.
\end{eqnarray}
Here and hereafter, the upper sign is chosen for $b>0$ and the lower one is chosen for $b<0$.

\subsection{$m \neq 0$}
In the case of $m \neq 0$, by substituting the deflection angle~$(\ref{eq:alpha_weak2})$ into  
the lens equation~(\ref{eq:Lens}) with Eq.~(\ref{eq:alpha02}), $n=0$, and $b=\theta D_{\mathrm{OL}}$,
reduced image angles $\hat{\theta} \equiv \theta/\theta_{\mathrm{E}0}$ are given by
\begin{eqnarray}
\hat{\theta}_{\pm0}(\hat{\phi})=\frac{1}{2}\left( \hat{\phi} \pm \sqrt{ \hat{\phi}^2+  4 } \right),
\end{eqnarray}
where $\hat{\phi} \equiv \phi/\theta_{\mathrm{E}0}$ is a reduced source angle  
and where $\theta_{\mathrm{E}0}$ is the Einstein ring angle given by
\begin{eqnarray}
\theta_{\mathrm{E}0}\equiv \theta_{+0}(0)= \sqrt{ \frac{4mD_{\mathrm{LS}}}{D_{\mathrm{OS}}D_{\mathrm{OL}}} }.
\end{eqnarray}
The magnifications of the image angles and its total magnification are given by
\begin{eqnarray}
\mu_{\pm0}
&\equiv& \frac{\hat{\theta}_{\pm0}}{\hat{\phi}} \frac{d\hat{\theta}_{\pm0}}{d \hat{\phi}} \nonumber\\
&=& \frac{1}{4} \left( 2 \pm \frac{\hat{\phi}}{\sqrt{\hat{\phi}^2+4}} \pm \frac{\sqrt{\hat{\phi}^2+4}}{\hat{\phi}} \right) \nonumber\\
&=& \frac{\hat{\theta}^4_{\pm 0}}{\left( \hat{\theta}^2_{\pm 0} \mp 1 \right) \left( \hat{\theta}^2_{\pm 0} \pm 1 \right)}
\end{eqnarray}
and  
\begin{eqnarray}
\mu_{0\mathrm{tot}}
&\equiv& \left| \mu_{+0} \right|+\left| \mu_{-0} \right|  \nonumber\\
&=& \frac{1}{2} \left( \frac{\hat{\phi}}{\sqrt{\hat{\phi}^2+4}} + \frac{\sqrt{\hat{\phi}^2+4}}{\hat{\phi}} \right),
\end{eqnarray}
respectively.

\subsection{$m = 0$}
In the case of $m = 0$, from the deflection angle~$(\ref{eq:alpha_weak2})$,  
the lens equation~(\ref{eq:Lens}), Eq.~(\ref{eq:alpha02}), $n=0$, and $b=\theta D_{\mathrm{OL}}$,
the magnifications of the image angles $\hat{\theta}_{\pm0}$ are expressed by
\begin{eqnarray}
\mu_{\pm0}
= \frac{\hat{\theta}^6_{\pm 0}}{\left( \hat{\theta}^3_{\pm 0} \mp 1 \right) \left( \hat{\theta}^3_{\pm 0} \pm 2 \right)}
\end{eqnarray}
and they can be calculated numerically by solving the lens equation. 
The Einstein ring angle $\theta_{\mathrm{E}0}$ is given by
\begin{eqnarray}
\theta_{\mathrm{E}0}= \left( \frac{\pi a^2D_{\mathrm{LS}}}{4 D_{\mathrm{OS}}D_{\mathrm{OL}}^2} \right)^{1/3}.
\end{eqnarray}

\section{Discussion and Conclusion}
In this paper, we have investigated gravitational lensing in the strong deflection limit in the Simpson-Visser spacetime. 
There are an antiphoton sphere on the throat and two photon spheres in a side and the other side of the throat for $2m\leq a < 3m$
while the antiphoton sphere and the throat coincide with the photon spheres and a marginally unstable photon sphere is formed at the throat for $a=3m$.
The deflection angle in the strong deflection limit has the form of Eq.~(\ref{eq:def0}) for $a \neq 3m$ and the form of Eq.~(\ref{eq:def02}) for $a = 3m$. 
In appendix~A, we will show that 
the Simpson-Visser spacetime for $a = 3m$ violates the assumptions of the strong deflection limit analysis for the the marginally unstable photon sphere
formed by the degeneracy of an antiphoton sphere and a photon sphere in Ref.~\cite{Tsukamoto:2020iez}.
This is similar to gravitational lensing in the strong deflection limit in a Damour-Solodukhin wormhole spacetime~\cite{Damour:2007ap} 
which has been investigated in Ref.~\cite{Tsukamoto:2020uay}.

We concentrate on only positive impact parameters or image angles in the strong deflection limit analysis.
However, the lens equation has negative solutions $\theta_{-n}\sim -\theta_n$ that represent negative image angles and every negative image angle makes a pair with the positive image angle.
The diameter of the pair images is given by $\theta_n-\theta_{-n} \sim 2\theta_{n}$.
Its magnification $\mu_{-n}$ of the image with $\theta_{-n}$ is obtained as $\mu_{-n} \sim -\mu_{n}$.

The parameters $\bar{a}$, $\bar{b}$, $\bar{c}$, and $\bar{d}$ of the deflection angles are shown in Table~I. 
We apply the strong deflection limit analysis into the supermassive object with its mass $m=m_*\equiv 4\times 10^6 M_{\odot}$ 
and its distant $D_{\mathrm{OL}}=8$~kpc at center of our galaxy and the observables in the strong deflection limit is shown in Table~I.
We notice that the size of the photon sphere do not depend on the value of $a/m$ for $a/m\leq 3$.
Therefore, we cannot distinguish the black hole from the wormhole 
if we assume Simpson-Visser metric and if there is no light source in the other side of the wormhole throat and 
if the mass, the distant, and the size of the photon sphere $2\theta_{\infty}$ were given.
It is very challenging future work to distinguish the black hole from the wormhole
by detecting the difference of the radii of the outermost relativistic ring and the photon sphere $\mathrm{s}\equiv \theta_{\mathrm{E}1}-\theta_\infty$. 
\begin{table*}[htbp]
 \caption{Parameters $\bar{a}$, $\bar{b}$, $\bar{c}$, and $\bar{d}$ of the deflection angle in the strong deflection limit and observables for given $a$ and $m$.
 The diameters of the innermost ring~$2\theta_{\infty}$, 
 the outermost ring among rings scattered by the photon sphere~$2\theta_{\mathrm{E}1}$, 
 the difference of the radii of the outermost ring and the innermost ring $\mathrm{s}=\theta_{\mathrm{E}1}-\theta_\infty$, 
 the magnification of the pair images of the outermost ring $\mu_{1\mathrm{tot}}(\phi) \sim 2 \left| \mu_{1} \right|$ for the source angle $\phi=1$ arcsecond, 
 and the ratio of the magnification of the outermost ring to the other rings $\mathrm{r}= \mu_1/\sum^\infty_{n=2} \mu_n$
 are shown in a case of $D_{\mathrm{OS}}=16$~kpc, $D_{\mathrm{OL}}=D_{\mathrm{LS}}=8$~kpc.
 $m_*$ and $a_*$ are defined as $m_*\equiv 4\times 10^6 M_{\odot}$ and $a_* \equiv 4(2/\pi)^{1/2} (D_{\mathrm{LS}}D_{\mathrm{OL}}/D_{\mathrm{OS}})^{1/4} m_*^{3/4}=7.2\times 10^{9}$km, respectively.
 Notice that we have defined $a_*$ so that the Ellis wormhole has the same diameter of the Einstein ring $2\theta_{\mathrm{E}0}=2.8618$~arcsecond as the ones in the cases of $m = m_*$.
 }
\begin{center}
\begin{tabular}{ c c c c c c c c c c c} \hline
$a$           	                   	    &$0$         &$m_*$        &$1.5m_*$    &$2m_*$       &$2.5m_*$    &$3m_*$      &$4m_*$       &$20m_*$    &$100m_*$    &$a_*$ \\ 
$m$           	                    	    &$m_*$       &$m_*$        &$m_*$       &$m_*$        &$m_*$       &$m_*$       &$m_*$        &$m_*$      &$m_*$       &$0$        \\ \hline
$\bar{a}$        	            	    &$1.0000$    &$1.0607$     &$1.1547$    &$1.3416$     &$1.8091$    &$\cdots$    &$2.0000$     &$1.0847$   &$1.0153$    &$1.0000$   \\ 
$\bar{b}$        	            	    &$-0.40023$  &$-0.46474$   &$-0.59088$  &$-0.92738$   &$-2.1867$   &$\cdots$    &$-1.6138$    &$-1.0031$  &$-1.0499$   &$-1.06215$ \\ 
$\bar{c}$    	                    	    &$\cdots$    &$\cdots$     &$\cdots$    &$\cdots$     &$\cdots$    &$3.13017$   &$\cdots$     &$\cdots$   &$\cdots$    &$\cdots$ \\ 
$\bar{d}$                           	    &$\cdots$    &$\cdots$     &$\cdots$    &$\cdots$     &$\cdots$    &$-2.74546$  &$\cdots$     &$\cdots$   &$\cdots$    &$\cdots$ \\ 
$2\theta_{\infty}$~[$\mu$as]        	    &$51.580$    &$51.580$     &$51.580$    &$51.580$     &$51.580$    &$51.580$    &$56.153$     &$209.27$   &$1002.7$    &$12028$ \\ 
$2\theta_{\mathrm{E}1}$~[$\mu$as]           &$51.645$    &$51.669$     &$51.714$    &$51.819$     &$52.058$    &$52.325$    &$57.236$     &$209.52$   &$1003.5$    &$12036$ \\ 
$\mathrm{s}$~[$\mu$as]                      &$0.032276$  &$0.044522$   &$0.066996$  &$0.11948$    &$0.23887$   &$0.37259$   &$0.54142$    &$0.12658$  &$0.36596$   &$3.8827$ \\ 
$\mu_{1\mathrm{tot}}(\phi) \times 10^{17}$  &$1.6163$    &$2.1029$     &$2.9093$    &$4.4747$     &$6.6649$    &$8.3751$    &$15.024$     &$23.709$   &$350.71$    &$45314$ \\ 
$\mathrm{r}$     	                    &$535.16$    &$373.40$     &$230.36$    &$107.62$     &$31.517$    &$11.2412$   &$22.550$     &$327.24$   &$486.47$    &$534.84$ \\ 
\hline
\end{tabular}
\end{center}
\end{table*}

We can estimate the value of $a/m$ from the observation of the photon sphere 
of the supermassive object at center of the giant elliptical galaxy M87.
Given mass $m=6.2 \times 10^9 M_{\odot}$ and $D_{\mathrm{OL}}=16.9$~Mpc, 
and if we assume that the diameter of the photon ring reported by Event Horizon Telescope Collaboration~\cite{Akiyama:2019cqa}
is equivalent to the diameter of the photon sphere $2\theta_{\infty}=42~\mu$as, 
we obtain $a/m \sim 4.2$.
Thus, the observation of the black hole shadow does not reject the wormhole with $a/m > 3.0$ which is disregarded in Refs.~\cite{Simpson:2018tsi,Nascimento:2020ime}.

\section*{Acknowledgements}
The author thanks an anonymous referee for his or her useful comments.
\appendix 
\section{Comparison with Tsukamoto~\cite{Tsukamoto:2020iez}}
The deflection angle in the strong deflection limit can be classified by $D_\mathrm{m}$ and its derivatives, 
where $D(r)$ is defined by 
\begin{eqnarray}
D(r)\equiv \frac{C^{\prime}}{C}-\frac{A^{\prime}}{A}.
\end{eqnarray}
A usual strong deflection limit analysis~\cite{Bozza:2002zj,Tsukamoto:2016jzh} for a photon sphere works only under assumptions 
$D_\mathrm{m}=0$ and $D^{\prime}_\mathrm{m}>0$
and 
a strong deflection limit analysis for a marginally unstable photon sphere investigated in Ref.~\cite{Tsukamoto:2020iez}
works under assumptions $D_\mathrm{m}=D^{\prime}_\mathrm{m}=0$ and $D^{\prime \prime}_\mathrm{m}>0$.

In the cases of $a<3m$ and $a>3m$, we obtain
\begin{eqnarray}
&&D_\mathrm{m}=0, \\
&&D^{\prime}_\mathrm{m}=-\frac{2(a^2-9m^2)}{27m^4}>0
\end{eqnarray}
and
\begin{eqnarray}
&&D_\mathrm{m}=0, \\
&&D^{\prime}_\mathrm{m}=\frac{2(a-3m)}{a^2(a-2m)}>0,
\end{eqnarray}
for the photon sphere at $r=r_\mathrm{m}=\sqrt{9m^2-a^2}$ and $r=r_\mathrm{m}=0$, respectively.
Therefore, we can apply the usual strong deflection limit analysis~\cite{Bozza:2002zj,Tsukamoto:2016jzh} to the cases. 

On the other hand, in the case of $a=3m$, we get
\begin{eqnarray}
&&D_\mathrm{m}=D^{\prime}_\mathrm{m}=D^{\prime \prime}_\mathrm{m}=0, \\
&&D^{\prime \prime \prime}_\mathrm{m}=\frac{2}{9m^4}>0,  
\end{eqnarray}
at the marginally unstable photon sphere at $r=r_\mathrm{m}=0$.
Therefore, we cannot apply the strong deflection limit analysis for the marginally unstable photon sphere in Ref.~\cite{Tsukamoto:2020iez} to the case of $a=3m$ 
since the assumption $D^{\prime \prime}_\mathrm{m}>0$ does not hold.



\end{document}